\documentstyle[prl,aps]{revtex}
\begin{document}
 \draft
\title{A Dynamical Approach to Temperature}
\author{Hans Henrik Rugh }
\address {Department of Mathematics, University of Warwick,
          Coventry, CV4 7AL, England}
\date{\today}
 \twocolumn[       
\maketitle
 \widetext       
 \begin{abstract}
We present a new dynamical approach for measuring the temperature of a
Hamiltonian dynamical system in the micro canonical ensemble
of thermodynamics.  We show that under the hypothesis of ergodicity
the temperature can be  computed as a time-average of the functional,
 $\nabla \cdot (\nabla H / \|\nabla H\|^2)$,
on the energy-surface.  Our method not only yields an efficient computational
approach for determining the temperature it also provides an intrinsic
link between dynamical systems theory and the statistical
mechanics of Hamiltonian systems.
 \end{abstract}
\pacs{05.20.Gg, 05.20.-y, 05.45.+b, 05.70.-a, 02.40.Vh, 02.40.-k}
 ] \narrowtext  

Two crucial assumptions allow one to pass
from the dynamical to the thermodynamical description
of a classical Hamiltonian system.
First, the system should exhibit some sort of ergodicity
so that time-averages can be replaced with equivalent
space-averages over the  micro-canonical
ensemble, i.e. over the corresponding energy surface
(possibly with other constraints) and
with respect to the invariant Liouville measure.
Second, it should make
sense to consider a large number of (weakly) coupled sub-systems
and look at the statistical behavior of an individual
fixed sub-system.
As the size of the ambient system tends to infinity
one obtains the thermodynamic limit in which 
the micro-canonical ensemble involving the  sub-system is assumed 
to behave like the canonical or the Gibbs ensemble where quantities like
temperature and pressure become operationally defined.
In a few special cases it has been possible to
prove the equivalence of the two ensembles but
it is believed  that under  quite general assumptions the two ensembles 
indeed give equivalent descriptions 
when systems are large. We refer to 
Ruelle  \cite{Ruelle} for a more complete treatment on the
equivalence of ensembles, to Lebowitz et al. \cite{LPV} for an illustrative
example of some differences between the ensembles
 and to Landau and Lifshitz \cite {LL} for a general
introduction to the subject.
For practical calculations using the micro canonical ensemble
see e.g. Evans and Morriss \cite{EM}.

In the thermodynamic description the temperature plays a
prominent role and the problem of measuring it using the
dynamical evolution of the system dates back to the early
days of thermodynamics. Already Maxwell realized that
when the Hamiltonian has the special form
$H(p,q,\ldots) = p^2/2 + K(q,\cdots)$, thus singling out
the square of a momenta,
 the canonical ensemble average of $p^2$ happens to be
the temperature. Thus assuming equivalence of ensembles and
ergodicity it suffices to measure the time-average of $p^2$ during
the dynamical evolution in order to get the temperature.

Here we shall pursue a different approach in which we focus
on the global geometric structure of the energy surface
We shall assume ergodicity only
and show that the  temperature as defined in the micro-canonical ensemble
can be obtained as a purely dynamical time-average of a
function which is related to the curvatures of the energy surface.

In this presentation we shall for clarity and
 without loss of generality restrict ourselves
to a Euclidean phase space\footnote{
     The Euclidean metric is not necessary but
     provides a convenient way of expressing
     the dynamical average 
     in terms of gradients of functions and 
     divergences of vector fields.},
 $\Omega = R^{2d}$\  $(d \geq 1)$,
and a Hamiltonian function, $H : \Omega \rightarrow R$, 
which preserves the Liouville measure
(here Lebesgue measure $m$) and such that for all 
energies considered 
 the set\ $H(\xi) \leq E$\ is bounded and has
a smooth and connected surface, $A(E) = \{H(\xi)=E\}$.
 Furthermore, we shall consider the case when
 the energy itself, $H=E$, is the only
preserved first integral. We refer to \cite{Rugh} for 
a generalization of this approach.

For the micro-canonical ensemble 
of classical thermodynamics one
defines the exponential of the entropy to be the 
(canonically invariant) weighted area
of an energy surface~:
\begin{equation}
  e^{S(E)} \equiv \int m(d\xi) \; \delta \left(
             H(\xi)\!-\!E \right) 
     = \int_{A(E)} 
 \frac{ m(dA_\xi)}{\|\nabla H(\xi)\|} ,
  \label{eq:entropy}
\end{equation}
and the temperature is obtained through differentiation
\begin{equation}
  \frac{1}{T(E)} = \frac{dS}{dE}  .
  \label{eq:temp}
\end{equation}
In practise a calculation of the entropy function
is impossible when the number
of degrees of freedom in the system is large.
However, we have the following~:\\

\noindent \underline{Theorem :}
{\em  Suppose that $(\Omega,H)$ is ergodic with respect
      to the Liouville measure $m_{|A(E)}$,
      restricted to a non-singular energy surface,
      $A(E)$. Then for 
      $m_{|A(E)}$-almost every $x(0)\in A(E)$, one has
  \begin{equation}
      \frac{1}{T(E)} = \lim_{t \rightarrow \infty} \frac{1}{t}
       \int_0^t d\tau\; \Phi(x(\tau))  ,
      \label{eq:dyntemp}
  \end{equation}
      where $x(\tau)$ is the trajectory of $x(0)$ under the 
      Hamiltonian flow and  the observable $\Phi$ equals  $\nabla \cdot
       (\nabla H / \|\nabla H\|^2)$,
      which can also be written as~:
  \begin{equation}
      \Phi = \frac{\Delta H}{\| \nabla H\|^2}
          - 2 \frac{D^2 H (\nabla H, \nabla H)}{\| \nabla H\|^4}  .
      \label{eq:Phi}
  \end{equation}
}

Remarks :\\[-3mm]
\begin{enumerate}
\item The measure and the energy are canonical invariants. 
      Hence, from equations (\ref{eq:entropy}) and (\ref{eq:temp}), so
      are the temperature and the time-average, equation (\ref{eq:dyntemp}),
      of $\Phi$ but {\em not} $\Phi$ itself.
      The theorem shows also that the temperature
      unlike the entropy is
      an intrinsic dynamical feature of the system
      (it can be computed as a dynamical average).
\item The theorem can quite easily, at the cost of making the notation
      more heavy, be generalized to arbitrary symplectic manifolds 
      without a Riemannian structure (cf. \cite{Rugh}).
\item It is of interest to see how the conjugated thermodynamic
      variables of other preserved first-integrals
      (e.g. total angular momentum) can be incorporated in the
      above dynamical formalism. It is for example well-known that
      the pressure can be obtained as a time-average of the
      momentum transfer per surface area element in the
      configuration space.
        It is also of interest to derive general
      \mbox{(thermo-)}dynamical relations for such quantities
      in the micro canonical ensemble (cf. \cite{Rugh}).
\item There is a formal resemblance of $\Phi$ to the Gauss
      curvature $1/R = \nabla \cdot (\nabla H / \|\nabla H\|)$,
      where for the latter quantity
      the norm of the gradient is not squared.
      In particular, when the number of degrees of freedom gets large
      (and the last term in the formula (\ref{eq:Phi}) becomes 
       relatively small) the relationship becomes closer.
      We do not know of any deeper reasons for this
      connection.
\end{enumerate}
     
The idea behind the above Theorem is to cast the energy dependence
of the geometry, i.e. the dependence on $A(E)$ in equations
(\ref{eq:entropy}) and (\ref{eq:temp}), into an algebraic
dependence for a fixed geometry. This 
is achieved through the introduction
of an auxiliary vector field (to a large extend arbitrary)
 and it will enable us to take the derivative
with respect to the energy inside the integral.

We consider in the following
two vector fields on $\Omega$~:
\begin{equation}
   v = J \nabla H \ \ \ \mbox{and} \ \ \ \eta = \frac{\nabla H}
       {\langle {\nabla H,\nabla H} \rangle}  .
\end{equation}
Here $J$ is a symplectic matrix, e.g.
 $\left(\begin{array}{rr}\bf{0} & \bf{1}\\-\bf{1}  & \bf{0} \end{array}
 \right)$, and
$\langle {\cdot,\cdot} \rangle$ is the usual Euclidean product in $\Omega$.
For a vector field $X$ on $\Omega$ we let $g^t_X$ denote the
associated flow. Thus $g^t_v$ is the Hamiltonian flow and
$g^s_\eta$ is the normalized energy gradient flow. Since
$\frac{d}{ds} H ( g^s_\eta(x)) = \langle{\nabla H(x),\eta(x)}\rangle = 1$
 we have 
\begin{equation}
 H ( g^s_\eta(x)) = s + H(x)  .
\end{equation}
Assuming that
no singular points are encountered
along the gradient flow, i.e. that $\nabla H \neq 0$ everywhere
on the energy surface (which is a  generic condition),
the map
\begin{equation}
  g^s_\eta : A(E) \rightarrow A(E+s)
  \label{eq:diffeo}
\end{equation}
is a diffeomorphism between energy surfaces.

Both 
the Liouville measure $m$
and the energy are  preserved under the Hamiltonian
flow, i.e.  $m(g^t_v dx)=m(dx)$ and $H(g^t_v(x))=H(x)$.
It follows that by 
considering an infinitesimal flow-box (under the energy gradient flow)
$g^{[0,\epsilon]}_\eta dA$ and
 by measuring the volume of such a flow-box we may define a measure $\mu$ on
the energy surface $A(E)$ which is also invariant under $g^t_v$,
\begin{equation}
   \mu(dA) = \lim_{\epsilon\rightarrow 0^+} \frac{1}{\epsilon}
             m \{ g^{[0,\epsilon]}_\eta dA \}  .
\end{equation}
For this measure, we have $e^{S(E)} = \int_{A(E)} \mu(dA)$.
The flow $g^s_\eta$ commutes with itself and hence it maps the  flow-box
$g^{[0,\epsilon]}_\eta dA(E)$ into
the flow-box $g^{[0,\epsilon]}_\eta dA(E+s)$ (by equation \ref{eq:diffeo}).
But then 
the ratios between the weighted
 areas of infinitesimal surface elements
just becomes a ratio between volume elements which
is nothing but the Jacobian of the
mapping~:
\[ \frac{\mu(dA(E+s))}{\mu(dA(E))} = 
          \det (\frac{ \partial g^s_{\eta}}{\partial x} )
          =  1 + s\; \nabla \cdot \eta + O(s^2)  \]
(to order $s$).
Defining $\Phi = \nabla \cdot \eta$ (i.e. our equation
\ref{eq:Phi}) we get the relation~:
\begin{equation}
 \int_{A(E+s)} \mu(dA) = \int_{A(E)} \mu(dA) (1+ s \Phi + ...)  ,
\end{equation}
and finally, using the definition of temperature (\ref{eq:temp}) we obtain~:
\begin{equation}
  \frac{1}{T(E)} = \frac{\int_{A(E)} \mu(dA) \Phi}
  {\int_{A(E)} \mu(dA)} .
\end{equation}
The interesting observation is now that if the Hamiltonian
flow is ergodic with respect to the Liouville measure
restricted to the energy surface,
$\mu$,
then by Birkhoff's theorem
 the right hand side equals the time-average (and thus equation
\ref{eq:dyntemp})
of $\Phi$ for $\mu$-almost every (equivalent to
$m_{|A(E)}$-a.e.)
 trajectory under the Hamiltonian flow,
thus proving our theorem.\\

An illustrative, though trivial\footnote
    {Strictly speaking this example does not meet the conditions
     of our Theorem as it is not ergodic.},
 example is given if we
 take $N$ harmonic
oscillators with the Hamiltonian
 $H = \sum_{i=1}^N \frac{1}{2}(q_i^2+p_i^2)$.
Its gradient
$\nabla H = (\vec{q},\vec{p})$ is non-vanishing at all non-zero
energies
and a straight forward
calculation yields $\Phi = (N-1)/{E} = {1}/{T(E)}$. In this particular
case we can directly read off
the micro-canonical result,
$E = (N-1)T$.
The fixation of the energy `freezes' one degree of freedom as is seen
by comparing with the  canonical result, $\langle E \rangle = NT$.
\\

In conclusion we have
 described a dynamical way of measuring
the temperature in the micro canonical ensemble of thermodynamics.
It is rigorously shown to work for a Hamiltonian system
at energies where the energy surface is regular and the
 flow is ergodic.
 The approach is
 purely dynamical and involves calculating the
time-average of the functional,
 $\nabla \cdot (\nabla H / \|\nabla H\|^2)$,
on the energy surface.
As the derivation shows
the dynamical temperature is really intrinsic to the
dynamical system in contrast to
the entropy itself. We note that in a classical theory
the entropy is in any case only defined up to a constant.
 We believe that the results and the underlying ideas
presented here may be of practical importance in numerical
simulations and helpful to physicists who wish
to understand the foundations of classical thermodynamics.

I am grateful to
 Predrag Cvitanovi\'c,
 Robert MacKay,
 David Ruelle as well as to many referees
for \mbox{(thermo-)}dynamic discussions.

\end{document}